\documentclass[aps,prb,twocolumn,amsmath,amssymb,showpacs,superscriptaddress]{revtex4-1}
\usepackage{graphicx}% Include figure files
\usepackage{dcolumn}% Align table columns on decimal point
\usepackage{bm}% bold math
\usepackage{times}
\usepackage{color}
\usepackage{aurical}
\usepackage[T1]{fontenc}
\usepackage[pdftex, colorlinks=true, linkcolor=blue, citecolor=blue,
urlcolor=blue]{hyperref}

\DeclareGraphicsExtensions{.pdf}

\newcommand{\br}{\mathbf{r}}

\newcommand{\dif}{\mathrm{d}}

\newcommand{\VT}{V_{\rm T}}

\newcommand{\calK}{\mathcal{K}}
\newcommand{\calL}{\mathcal{L}}
\newcommand{\calD}{\mathcal{D}}
\newcommand{\calB}{\mathcal{B}}

\def\ba{\bar{a}}

\def\Me{M_{\rm e}}
\def\coola{\mbox{\Fontauri A}} % the fancy font for the little a matrix elements
\newcommand{\saclay}{Service de Physique de l'\'{E}tat Condens\'{e}, 
                     CNRS URA 2464, CEA Saclay, 
                     F-91191 Gif-sur-Yvette, France}
\newcommand{\strasbourg}{Institut de Physique et Chimie des Mat\'{e}riaux 
                         de Strasbourg, 
                         Universit\'{e} de Strasbourg, 
                         CNRS UMR 7504, 
                         F-67034 Strasbourg, France}

\begin{document}

\title{Scanning-gate-induced effects in nonlinear transport through nanostructures} 

\author{Cosimo Gorini}
\affiliation{\saclay}
\affiliation{\strasbourg}
%\affiliation{\augsburg}
\author{Dietmar Weinmann}
\affiliation{\strasbourg}
\author{Rodolfo A. Jalabert}
\affiliation{\strasbourg}

%\date{\today}

\begin{abstract}
We investigate the effect of a scanning gate tip on the nonlinear 
quantum transport properties of nanostructures. Generally, we predict that the 
symmetry of the current-voltage characteristic in reflection-symmetric samples 
is broken by a tip-induced rectifying conductance correction. 
Moreover, in the case of a quantum point contact (QPC), the tip-induced 
rectification term becomes dominant as compared to the change of the linear 
conductance at large tip-QPC distances. 
Calculations for a weak tip probing a QPC modeled by an abrupt 
constriction show that these effects are experimentally observable.
\end{abstract}

\pacs{72.10.-d, %Theory of electronic transport; scattering mechanisms
      73.23.-b, %Electronic transport in mesoscopic systems
      07.79.-v, %Scanning probe microscopes and components
      72.20.Ht  %High-field and nonlinear effects 
      }

\maketitle

%%%%%%%%%%%%%%%%%%%%%%%%%%%%%%%%%%%%%%%%%%%%%%%%%%%%%%%%%%%%%%%%%%%%%%%%%%%%%%%

\section{Introduction}

Nonlinear transport in semiconductor devices is the most common situation, 
but the analysis is considerably more complicated than in the linear case. 
While in the linear response regime a knowledge of the actual electric field 
distribution is not required to obtain the dissipation in the system, 
the field distribution does matter for many applications beyond linear 
transport.\cite{vanhouten1992,landauer1987,landauer1989} 
Thus, a great difficulty facing nonlinear transport theories is the necessity 
to consider the self-consistent potential $\phi({\br})$ resulting from 
the imposed voltages between the probes and the electron-electron interactions 
in the device \cite{buttiker1993,christen1996a} (i.e., the self-gating effect). 

The use of a Scanning Tunneling Microscope (STM) to obtain information about 
the local field  was proposed 25 years ago, 
\cite{kirtley1988,landauer1989,vanhouten1992} but only recently 
\cite{jura2009,jura2010,kozikov_unpub,sellier2013} a related technique, 
the Scanning Gate Microscopy (SGM), has been applied in the nonlinear regime 
to study electron-electron scattering in a two-dimensional electron gas (2DEG) 
surrounding a Quantum Point Contact (QPC). The SGM appears as a less invasive 
probe than the STM, as it consists of a charged atomic force microscope scanning 
over the sample and thus modifying the conductance only through a capacitive 
coupling to the buried 2DEG.\cite{binnig1987, giessibl2003, sellier2011}  

The recent works of SGM in the nonlinear regime have been preceded by an 
important activity in the study of the tip-induced changes of the linear 
conductance through a 
QPC \cite{topinka2000, topinka2001, leroy2005, heller2005, schnez2011a} 
and other mesoscopic systems.\cite{martins2007,schnez2011a,kozikov2013} 
Even in the linear regime, the interpretation of the resulting scans 
is delicate.\cite{jalabert2010, gorini2013}
On one hand, from various experimental and theoretical works focused on 
a QPC probed by a strongly charged tip, the conductance change appears 
to be closely related to the local current 
density.\cite{topinka2001, leroy2005, metalidis2005, cresti2006} 
On the other, it has been shown \cite{gorini2013} that only under quite 
restrictive conditions (a spatially symmetric QPC tuned to a conductance 
plateau) the tip-induced conductance change is directly related to the 
current density at the tip position.

In the nonlinear regime the SGM of a QPC has delivered some intriguing results. 
The tip-induced conductance correction appears asymmetric in the bias voltage 
$V$ and reverses its sign for large $V$. \cite{jura2010} 
While an interpretation in terms of the nonequilibrium distribution of electrons 
in a localized region of the 2DEG near the QPC was proposed, further experimental 
and theoretical work appeared necessary in order to justify the use of an 
effective electron temperature.\cite{jura2010} Working in the regime of a partially
closed QPC, an oscillatory splitting of the zero-bias anomaly with tip position, 
correlated with simultaneous appearances of the 0.7 anomaly, has been recently 
reported.\cite{sellier2013} 
These findings concerning an SGM setup in the regime of nonlinear transport 
through a QPC illustrate the need to address two related questions. 
Firstly, which is the local potential of a QPC operating 
in the nonlinear regime? 
\cite{vanhouten1992,pepper1992,ouchterlony1995,kristensen2000,gloos2006,
song2009}
Secondly, what is actually measured in the scanning gate microscopy of a 
QPC in the linear and nonlinear 
regimes?\cite{jalabert2010, sellier2011, gorini2013, schnez2011a}  

In this work we provide a theoretical approach to the SGM of a QPC operating 
in the nonlinear regime, by suitably generalizing the linear response approach of
Refs.~\onlinecite{jalabert2010, gorini2013} within the general gauge-invariant
framework defined in Ref.~\onlinecite{christen1996a}.
That is, in order to keep the problem tractable and to stay on a rigorous basis we limit 
ourselves to a gauge-invariant theory of weakly nonlinear transport, using a 
one-particle scattering approach and a perturbative tip.  We underline the 
asymmetries appearing in nonlinear transport and predict two qualitative effects: 
(i) an odd-in-bias conductance correction induced by the tip in a nominally 
symmetric QPC; (ii) for increasing tip-QPC distances, a slower
decay of the nonlinear conductance corrections as compared to the linear one. 
We investigate the quantitative behavior of the nonlinear conductance 
by solving the special case of an abrupt QPC subject to a finite bias. 
Recent experiments have shown that almost ideal, perfectly symmetric QPCs 
can be realized.\cite{roessler2011}  Tip-induced asymmetries should be 
observable in such systems and provide a signature of the probe's invasiveness.

\section{Nonlinear transport coefficients}

\begin{figure}
\includegraphics[width=\linewidth]{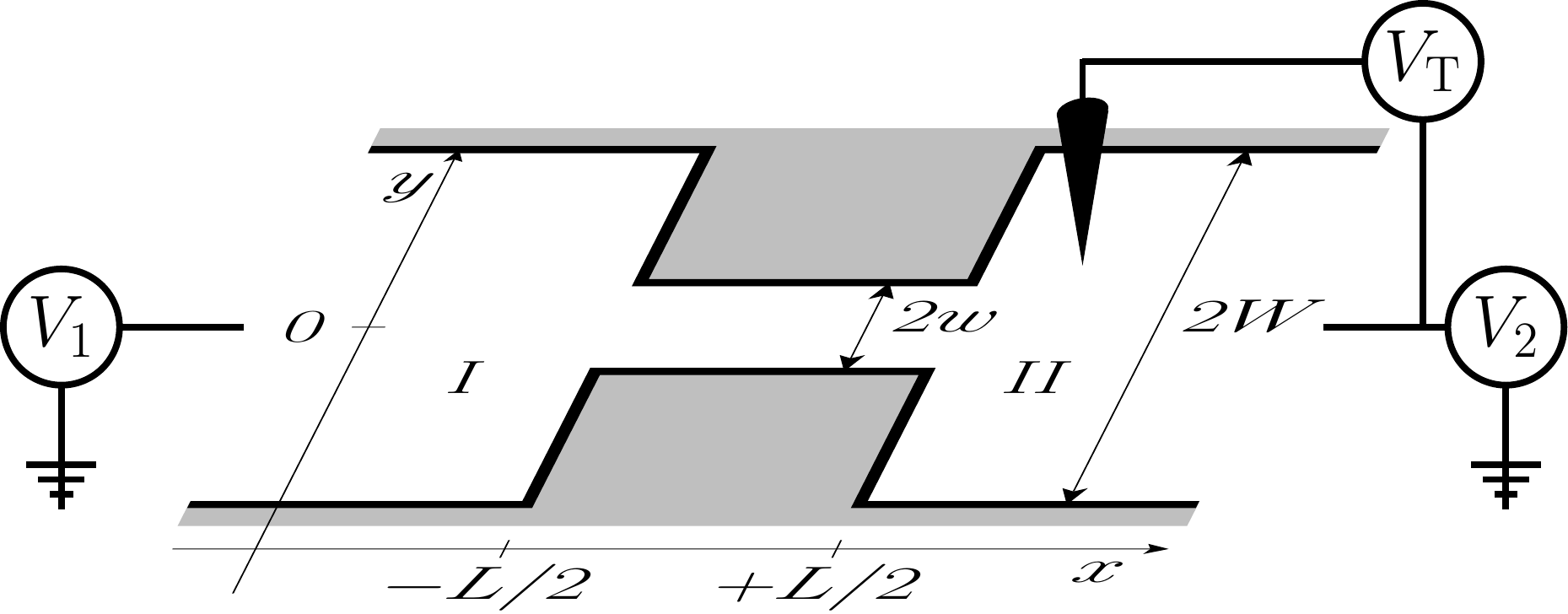}
\caption{\label{fig1}
Sketch of the considered setup. A QPC in a two-dimensional electron gas is connected 
via wide leads to voltage sources generating voltages $V_1$ and $V_2$ in the left and right 
reservoir, respectively. An SGM tip acting in region II at the right of the constriction 
is at a voltage $V_\mathrm{T}$ with respect to $V_2$. For quantitative purposes, 
we use an abrupt geometry with a narrow region of width $2w$ and length $L$ between leads 
of width $2W$. 
}
\end{figure}
In the two-terminal configuration sketched in Fig.\ \ref{fig1} the voltage $V_1$ ($V_2$) is 
imposed at the left (right) reservoir and the tip acting at the right of the QPC (region II) is at 
$V_\mathrm{T}$ with respect to $V_2$. While the depicted abrupt QPC is the example we use to calculate 
quantitative results, the general results that we will present are valid in any 
phase-coherent device. Gauge invariance implies that the measurable quantities do not change 
upon an overall shift of the energies of the problem. Thus, the current $I$ through the device 
does not depend on the reference voltage $U = (V_1 + V_2)/2$, but only on the bias voltage 
$V = V_1 - V_2$ according to
\begin{equation}
\label{ouverture}
I(V) = \frac{2e^2}{h} \left( g_1 \  V + \frac{1}{2} \ g_2 \ V^2 
+ \frac{1}{3!} \ g_3 \ V^3+\mathcal{O}(V^4) \right) \, .
\end{equation}
Scaling out the conductance quantum $2e^2/h$ allows us to work with the dimensionless differential conductance
$g(V)=(h/2e^2)\partial I/\partial V$, depending on the dimensionless linear,
second-, and third-order conductances, $g_1$, $g_2$, and 
$g_3$, respectively. 
Within the general approach of Ref.~\onlinecite{christen1996a}, for a two-terminal device 
operating at a low temperature $T$ 
(in the limit $k_\mathrm{B}T\ll\epsilon_\mathrm{F}$, with $k_\mathrm{B}$ 
the Boltzmann constant and $\epsilon_\mathrm{F}$ the Fermi energy) 
the key quantity describing electron transport is the screened transmission probability 
$\mathcal{T}(\varepsilon,\left\{V_1, V_2\right\})=\mathrm{Tr}[t^\dagger t]$ 
depending on the energy $\varepsilon$ of the transmitted electron 
and on the applied voltages [through the self-consistent potential $\phi(\br)$]. 

We use the standard notation of $t(t')$ and $r(r')$ for the transmission and 
reflection submatrices of the scattering matrix for particles impinging from the 
left (right) side of the scatterer and write
\begin{subequations}\label{eq:coefficients} 
\begin{eqnarray}
g_1 &=& \mathcal{T} \, , \\
g_2 &=& (\partial_{V_1}-\partial_{V_2})\mathcal{T} 
= 2\partial_V\mathcal{T} \, , 
\\
g_3 &=& (\partial_{V_1}\partial_{V_1}-\partial_{V_1}\partial_{V_2}
        +\partial_{V_2}\partial_{V_2})\mathcal{T} 
\nonumber \\
&=& [3\partial_V\partial_V+(1/4)\partial_U\partial_U]\mathcal{T} \, .
\end{eqnarray}
\end{subequations}
The second equalities follow from gauge invariance. 
The expressions involving $U$ and $V$-derivatives should be evaluated at 
${(\epsilon_\mathrm{F}, \left\{V=0,U=0\right\})}$, 
while the others at ${(\epsilon_\mathrm{F}, \left\{V_1=V_2=0\right\})}$. 

\begin{figure}
\includegraphics[width=\linewidth]{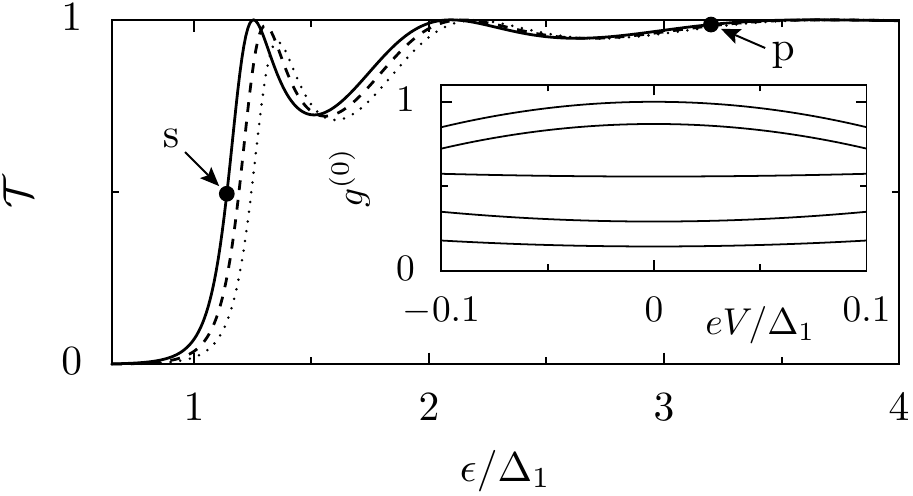}
\caption{\label{fig2}
Unperturbed transmission probability $\mathcal{T}$ through an abrupt QPC with 
$L/w=2.5$ (see Fig.\ \ref{fig1}) as a function of the energy $\epsilon$. 
Solid, dashed, and dotted lines are evaluated from Eq.\ \eqref{eq:tm} below for
voltages $eV/\Delta_1=0$, $0.1$, and $0.2$, respectively, where $\Delta_1$ is the energy of the 
first quantized transverse mode in the constriction.
The points ``s'' and ``p'' mark the positions in the step and on the plateau where 
Figs.~\ref{fig3} and \ref{fig4} are evaluated, respectively. 
The inset shows the voltage dependence of the differential conductance, without tip, 
up to third order for different Fermi energies at the first step.
}
\end{figure}
Starting without the SGM tip, we present in Fig.\ \ref{fig2} the unperturbed 
transmission probability $\mathcal{T}$ as a function of $\epsilon$ for various 
bias voltages $V$, evaluated for the QPC sketched in Fig.\ \ref{fig1}. 
The unperturbed differential conductance up to third order 
$g^{(0)}=g^{(0)}_1+g^{(0)}_2V+(g^{(0)}_3/2)V^2$ is shown in the inset of 
Fig.\ \ref{fig2}. The $g_i^{(0)}$ are given by Eq. \eqref{eq:coefficients} when 
using as $\mathcal{T}$ the tip-unperturbed transition probability (to simplify 
the notation we do not write the index $(0)$ in $\mathcal{T}$ or in the 
scattering submatrices).
The energy-dependent features in the transmission and conductance plateaus 
characteristic of clean abrupt geometries have been shown to be smoothed by 
finite temperature and bias.\cite{szafer1989,lindelof08} 
The width of the conductance plateaus is considerably reduced only for 
rather large bias voltages \cite{kouwenhoven1989} ($V\gtrsim \Delta_1/e$), 
where $\Delta_1$ is the lowest transverse energy in the constriction. 

\section{Scanning-gate effects on transport coefficients}

We now consider the action of an SGM tip. The voltage $V_\mathrm{T}$ is applied 
with respect to the reference $V_2$ in order to render the former gauge invariant.
In an SGM setup the linear, second, and third-order conductances of the 
unperturbed device will change under the effect of a perturbing voltage $V_\mathrm{T}$.
According to Ref.~\onlinecite{jalabert2010}, the tip-induced changes in the 
conductance coefficients $g_i^{(1)}$ are obtained when $\mathcal{T}$ in 
Eq.\ \eqref{eq:coefficients} is replaced by  
\begin{eqnarray}
\label{eq:dgs_ugly}
\mu(\varepsilon,\left\{V_1, V_2\right\}) &=&
      -4 \pi\ \mathrm{Im}\left\lbrace \mathrm{Tr}
      \left[r^{\dagger}t'\ \mathcal{V}^{21}\right]\right\rbrace
      \nonumber
\\  \label{eq:dgs_nice}
    &=&-4\pi\ \mathrm{Im}\left\{\sum_{m=1}^N
      r_m^*t'_m\mathcal{U}^{21}_{mm}\right\} \ .
\end{eqnarray}
The matrix elements of the perturbing potential $\VT(\br)$ in the basis 
of the scattering states $\Psi_{l,\varepsilon,a}$ are 
\begin{equation}
\label{eq:V21}
\mathcal{V}^{\bar{l}l}_{\ba a} =
\int \dif \br \
\Psi_{\bar{l},{\varepsilon},\bar{a}}^{*}(\br) \ \VT(\br) \ 
\Psi_{l,\varepsilon,a}(\br), 
\end{equation}
where $l$ and $a$ represent the lead and mode, respectively, from which 
the scattering state (with energy $\varepsilon$) impinges. The scattering 
submatrices and the scattering states are those of the bias-dependent, 
tip-unperturbed problem. 
The last equality of Eq.~\eqref{eq:dgs_nice} is obtained by a change into 
the basis of scattering eigenstates 
(built from the eigenmodes of $t^\dagger t$), \cite{gorini2013} 
where the reflection and transmission submatrices are diagonal 
with non-zero elements $r_m$ ($r'_m$) and $t_m$ ($t'_m$) for $l=1,2$.
$N$ is the number of channels in the leads and $\mathcal{U}^{21}_{m'm}$ 
is the matrix element of the perturbing potential between the 
$m'$\textsuperscript{th} left- and the $m$\textsuperscript{th} 
right-moving scattering eigenstates.

For a QPC $r_m t'_m = 0$ on a conductance plateau,\cite{jalabert2010,gorini2013} 
yielding a vanishing $\mu$. Moreover, away from the edges of the plateau the 
$V$ and $U$-derivatives of $r_m t'_m$ also vanish, and thus 
$g_i^{(1)}=0$. Hence, the tip-induced changes in the linear, 
second, and third-order conductances are dominated by 
$g_i^{(2)}$. Those corrections scale as $V_\mathrm{T}^2$ 
and are obtained when $\mathcal{T}$ in Eq.\ \eqref{eq:coefficients} 
is replaced by  
\begin{equation}
%\label{eq:dgp_ugly}
\nu(\varepsilon,\left\{V_1, V_2\right\}) 
%=- 4 \pi^2\ \mathrm{Tr}
%      \left[t^{\dagger}t\ \mathcal{V}^{12}\mathcal{V}^{21}\right]
%      \nonumber
%\\  
\label{eq:dgp_nice}    
    =-4\pi^2 \sum_{m,m'=1}^M
       \mathcal{U}^{12}_{mm'}\mathcal{U}^{21}_{m'm},
\end{equation}
where $M$ is the number of open channels in the constriction.  

\section{Tip-induced symmetry breaking}

Various properties of the above discussed conductances, to different orders 
in $V$ and $V_\mathrm{T}$, can be studied depending on the characteristics of 
the QPC and the regime of operation. Interestingly, general properties 
can be inferred from symmetry considerations. Onsager's relations for 
linear response \cite{onsager1931,casimir1945,buttiker1986} and their 
generalization to the nonlinear regime \cite{lofgren2004,sanchez2004,andreev2006} 
determine the symmetry of the response functions. 
For a left-right symmetric device in the absence of a magnetic field, 
the $I$-$V$ characteristics is odd, i.e. $I(-V)=-I(V)$ and $g_2^{(0)}=0$. This is the reason 
for the symmetry observed in the inset of Fig.\ \ref{fig2}.
When a symmetric QPC is approached by a perturbing tip, the 
spatial symmetry is broken and one expects $g_2^{(1)}\neq 0$ at a conductance 
step and $g_2^{(2)}\neq 0$ on a conductance plateau. The tip-induced 
second-order conductance is a rectification effect, observable in 
nominally symmetric devices.

In order to quantify the above described effect one needs to solve the 
scattering problem with a finite bias, which requires modeling the 
constriction and the self-gating effect. The saddle-point model, 
applicable to smooth and relatively short QPCs, was the basis of 
numerous studies in the nonlinear regime,\cite{pepper1992,ouchterlony1995}
and the close comparison with experiments allows to extract 
the constriction's geometrical parameters.\cite{gloos2006,song2009} 
A symmetric potential drop between the reservoir and the bottleneck 
is compatible with experimental results.\cite{pepper1992,kristensen2000}
The saddle-point model is appropriate for studies of 
the unperturbed conductance, determined by the features of the region 
immediately surrounding its narrowest point.
However the tip-dependent conductance changes depend on the wave-functions 
far away from the bottleneck, where the saddle-point model does not provide 
a good description. 
This is why for an unbiased abrupt constriction, describing a hard-wall 
and relatively long QPC, a generalization of the mean-field 
approximation \cite{szafer1989} was developed 
to obtain the scattering eigenstates.\cite{gorini2013} 

In the biased case we assume the electric field to be non-zero only 
in the constriction itself.
Such an assumption is supported by theoretical calculations showing that 
the potential drop for diffusive and ballistic constrictions occurs 
in the vicinity of the contact at distances of up to the order of the 
contact size,\cite{vanhouten1992,rokni1995} and has been used in 
numerical approaches yielding a reasonable account of weak nonlinear 
effects in abrupt QPCs.\cite{casta1990} Since we do not describe the 
physics of strong bias and half-plateaus,\cite{glazman1989} but we 
only consider weak nonlinearities, assuming a linear potential drop 
between $V_1$ and $V_2$ within the constriction, without inelastic 
effects, is appropriate. In a symmetric QPC the potential drop does not 
have a quadratic component.  
Calculations up to $g_3$ are therefore consistent with our assumptions.

\section{Application to an abrupt quantum point contact}

We consider an abrupt QPC (see Fig.\ \ref{fig1}) with 
hard wall boundaries confining the electrons to a narrow strip of length 
$L$ and width $2w$ in the central region, being directly attached to 
leads of width $2W$. The transverse channel wavefunctions are
$\phi_a(y)=((-1)^p/\sqrt{W})\sin\left[q_a(y-W)\right]$, with $q_a=\pi a/2W$ 
and $p={\rm Int}\{a/2\}$. 
The outgoing $(+)$ and ingoing $(-)$ modes for left ($l=1$) 
and right ($l=2$) leads read
\begin{equation}\label{eq:modes}
\varphi_{l\varepsilon a}^{(\pm)}(\br)=
\frac{c}{\sqrt{k_{l a}}} \ e^{[\pm(-1)^l ik_{l a}x]} \ \phi_a(y)
\, ,
\end{equation}
with $\br=(x,y)$, $c=\sqrt{\Me/2\pi\hbar^2}$, and 
longitudinal wavevectors satisfying 
$k_{l a}^2=k_{l}^2-q_a^2$. Here 
$k_{l} = \sqrt{(2\Me/\hbar^2)(\varepsilon-eV_{l})}$, while
$e$ and $\Me$ stand for the charge and the effective mass of the electrons. 
The important difference with the linear case is that for a given energy 
$\varepsilon$ the longitudinal wave-vector $k_{l a}$ differs 
at the two extremes of the junction according to the imposed voltages 
$V_1$ and $V_2$.
Moreover, in the central region the scattering wave-function for electrons 
impinging from mode $a$ in lead $l$ is expanded as 
\begin{equation}
\Psi_{l\varepsilon a}(\br)= c \sum_{n=1}^{\infty} 
\left[\gamma^+_{l na}f_n(x)+\gamma^-_{l na}g_n(x)\right]\Phi_n(y)
\end{equation}
with $\Phi_n(y)=(1/\sqrt{w})\sin\left[Q_n(y-w)\right]$ the transverse 
wavefunctions in the narrow region ($Q_n=n\pi/2w$), while 
$f_n(x)$ and $g_n(x)$ are the two Airy functions 
resulting from our assumption of a linear potential $\phi(\br)$ 
within the constriction. 
The overlaps of the transverse channel wavefunctions 
\begin{equation}\label{coola}
\coola_{na} = \int_{-w}^{w}\dif y\,\Phi_{n}(y)\phi_a(y) \, ,
\end{equation}
together with the momentum-like quantity
\begin{equation}
\calK_{l n n'}={\sum_a}' k_{l a} \ \coola_{na}\coola_{n'a} \, ,
\end{equation}
play a key role in the solution of the linear system of equations arising 
from the wave-function matching at $x=\pm L/2$.\cite{gorini2013,szafer1989} 
(Here, ${\sum'_a}$ denotes the sum over modes $a$ with 
the same parity as $n$ only). 
Since the $\coola_{na}$'s are appreciably different from zero 
only for $q_a \in [Q_{n-1},Q_{n+1}]$ and $k_a$ is a smooth function of $q_a$, 
one has $\mathcal{K}_{l nn'} \approx\mathcal{K}_{l n}\delta_{nn'}$. 
The above-cited approximations lead to the scattering amplitudes
\begin{widetext}
\begin{subequations}
\label{eq:scattampl}
\begin{eqnarray}
t_{ba}&=&2i {\mathcal W} \ \sqrt{k_{2b}k_{1a}} \ 
\exp{\left[-i\left(k_{2b}+k_{1a}\right)\frac{L}{2}\right]} \
\sum_n
\frac{\coola_{nb}\coola_{na}}{\calD_n}  \, ,
\\
r_{ba}&=& -\delta_{ba} \exp{[-ik_{1b}L]}
+2i {\mathcal W} \ \sqrt{k_{1b}k_{1a}} \ 
\exp{\left[-i\left(k_{1b}+k_{1a}\right)\frac{L}{2}\right]} \
\sum_n 
\frac{\calB_{1n}\coola_{nb}\coola_{na}}{\calD_n}  \
 \, ,
 \\
 t_{ba}^{\prime}&=&2i {\mathcal W} \ \sqrt{k_{1b}k_{2a}} \ 
\exp{\left[-i\left(k_{1b}+k_{2a}\right)\frac{L}{2}\right]} \
\sum_n
\frac{\coola_{nb}\coola_{na}}{\calD_n}  \, ,
\\
r_{ba}^{\prime}&=& -\delta_{ba} \exp{[-ik_{2b}L]}
+2i {\mathcal W} \ \sqrt{k_{2b}k_{2a}} \ 
\exp{\left[-i\left(k_{2b}+k_{2a}\right)\frac{L}{2}\right]} \
\sum_n 
\frac{\calB_{2n}\coola_{nb}\coola_{na}}{\calD_n}  \
 \, ,
\end{eqnarray}
\end{subequations}
with the definitions
\begin{subequations}
\label{eq:definitions}
%\begin{eqnarray*}
\begin{eqnarray}
\calD_{n} &=& [\calL^+_{1n}g_n(-L/2)][\calL^-_{2n}f_n(L/2)]
              -[\calL^+_{1n}f_n(-L/2)][\calL^-_{2n}g_n(L/2)] \, ,
\\
\calB_{1(2)n} &=& g_n(\mp L/2)\calL^{\mp}_{2(1)n}f_n(\pm L/2)
                 -f_n(\mp L/2)\calL^{\mp}_{2(1)n}g_n(\pm L/2)\, ,
%\end{eqnarray*}
\end{eqnarray}
\end{subequations}
where $\calL^{\pm}_{l n}=\partial_x \pm i\calK_{l n}$. 
The Wronskian of $f_n$ and $g_n$ is the constant
${\mathcal W}=-(1/\pi)[4\pi c^2 eV/L]^{1/3}$ . 

The solution \eqref{eq:scattampl} allows us to build the scattering eigenstates 
$\chi_{l\varepsilon m}$  as superposition of the scattering states 
$\Psi_{l\varepsilon a}$. 
The corresponding wave-function in the wide regions I and II, 
for a mode impinging from the left, can be asymptotically expressed as
\begin{subequations}
\label{eq:scatteringeigenmodes_1} 
\begin{eqnarray}
\chi^\mathrm{I}_{1\varepsilon m}(\br)  
&=&   \frac{c}{\sqrt{k_{1}}}\sqrt{\frac{2}{\pi\rho w}}
      \frac{\Theta_m^\mathrm{I}(\theta)}{\sqrt{\mathrm{Re}\{\calK_{1m}\}}}
      \left\{e^{-i(k_{1}\rho-\pi/4)}+r_m
             e^{i(k_{1}\rho-\pi/4)}\right\}  \, ,
\\
      \chi^\mathrm{II}_{1\epsilon m}(\br) 
&=& 
      \frac{c}{\sqrt{k_{2}}}\sqrt{\frac{2}{\pi\rho w}}
      \frac{\Theta_m^\mathrm{II}(\theta)}{\sqrt{\mathrm{Re}\{\calK_{2m}\}}}      
      \ t_m e^{i(k_{2}\rho-\pi/4)}  \, .
\end{eqnarray}
\end{subequations}
For a mode impinging from the right, we have 
\begin{subequations}
\label{eq:scatteringeigenmodes_2} 
\begin{eqnarray}
\chi^\mathrm{II}_{2\varepsilon m}(\br)  
&=&   \frac{c}{\sqrt{k_{2}}}\sqrt{\frac{2}{\pi\rho w}}
      \frac{\Theta_m^\mathrm{II}(\theta)}{\sqrt{\mathrm{Re}\{\calK_{2m}\}}}
      \left\{e^{i(k_{2}\rho-\pi/4)}+r_m^{\prime}
             e^{-i(k_{2}\rho-\pi/4)}\right\}  \, ,
\\
      \chi^\mathrm{I}_{2\epsilon m}(\br) 
&=& 
      \frac{c}{\sqrt{k_{1}}}\sqrt{\frac{2}{\pi\rho w}}
      \frac{\Theta_m^\mathrm{I}(\theta)}{\sqrt{\mathrm{Re}\{\calK_{1m}\}}}      
      \ t_m^{\prime}e^{-i(k_{1}\rho-\pi/4)}  \, .
\end{eqnarray}
\end{subequations}
\end{widetext}
We denote $(\rho, \theta )$ the polar coordinates of $\br$ in a system 
centered at the entrance (exit) of the constriction when $\br$ is in 
region I (II). The angular dependence of the wavefunctions is given by 
\begin{subequations} 
\begin{eqnarray}
\Theta_m^\mathrm{I}(\theta) &=& (-1)^{m} \ \frac{Q_m k_{1} \cos\theta \, 
  f_m(k_{1}w\sin\theta)}{(k_{1}\sin\theta)^2-Q_m^2}  \, ,
\\
\Theta_m^\mathrm{II}(\theta) &=&  \frac{
      Q_m k_{2} \cos\theta \, f_m(k_{2}w\sin\theta)}
           {(k_{2}\sin\theta)^2-Q_m^2} 
           \, ,
\end{eqnarray}
\end{subequations}
with $f_m(z)=-[e^{iz}-(-1)^{m} e^{-iz}]/2$. The transmission and reflection 
amplitudes associated with the scattering eignemodes are 
\begin{subequations}
\begin{eqnarray} \label{eq:tm}
t_m &=& 
t_m^{\prime} = 2i \ \frac{\mathcal W}{\calD_m} 
\sqrt{\mathrm{Re}\{\calK_{1m}\} \ \mathrm{Re}\{\calK_{2m}\}}  \, \\
r_m &=& 
\frac{2i\mathrm{Re}\{\calK_{1m}\} \ \calB_{1m}}{\calD_m}-1 \, ,
\\
r_m^{\prime} &=& 
\frac{2i\mathrm{Re}\{\calK_{2m}\} \ \calB_{2m}}{\calD_m}-1 \, ,
\end{eqnarray}
\end{subequations} 

\begin{figure}
\includegraphics[width=\linewidth]{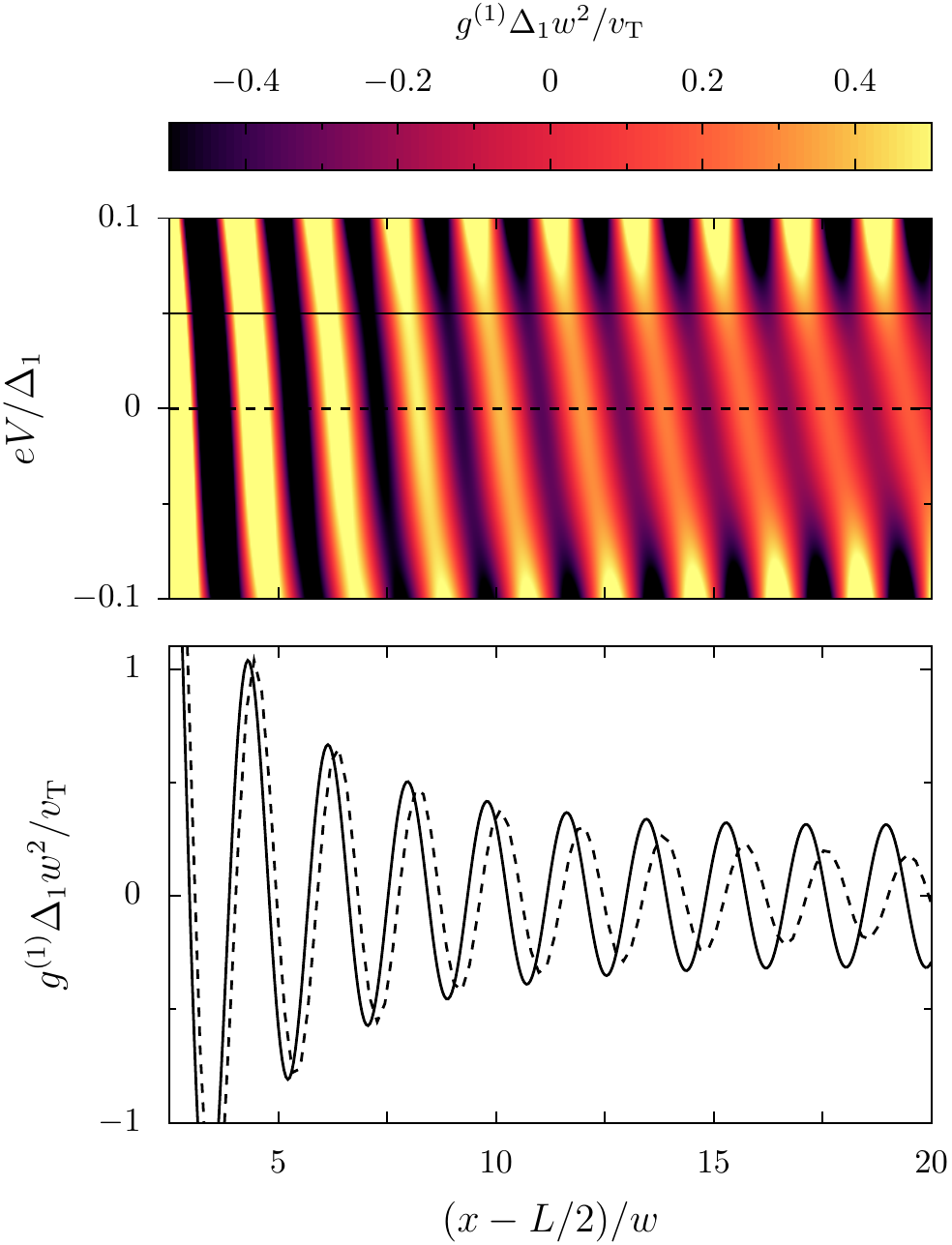}
\caption{\label{fig3}
Upper panel:
Colorscale plot of the tip-induced change of the differential conductance up to 
third-order terms, $g^{(1)}(V)=g_1^{(1)}+g_2^{(1)}V+g_3^{(1)}V^2/2$, as a function of 
bias voltage $V$ and tip position along the $x$-axis at the first conductance step 
($\lambda_{\rm F}\approx4 w$, point ``s'' in Fig.\ \ref{fig2}) 
of the abrupt QPC sketched in Fig.~\ref{fig1} ($L/w=2.5$). 
The conductance correction is obtained to first order in a local tip potential. 
The tilt of the oscillating pattern in the upper panel at large tip-QPC distances 
is a signature of the tip-induced $V$-asymmetry.
Lower panel: Dashed and solid lines correspond to the cuts indicated at $V=0$ and 
$eV=0.05\Delta_1$ in the upper panel.
}
\end{figure}

From \eqref{eq:tm} we get the transmission probability without the 
tip $\mathcal{T}$. Its energy dependence is shown in 
Fig.\ \ref{fig2} for different values of the bias.  
From the expressions 
\eqref{eq:scatteringeigenmodes_1}-\eqref{eq:scatteringeigenmodes_2} 
of the scattering eigenstates, and given the tip potential 
$V_\mathrm{T}(\br)$, we obtain the coefficients $\mu$, and $\nu$. 
We thus have closed expressions for the linear, 
second, and third-order conductances, as well as their tip-induced corrections. 
Fig.\ \ref{fig3} presents the change of the differential conductance when an 
SGM tip scans the $x$-axis of an abrupt QPC tuned to the first conductance step 
(point ``s'' in Fig.\ref{fig2}) for the case of a local tip potential 
$V_\mathrm{T}(\br)=v_\mathrm{T}\delta(\br-\br_\mathrm{T})$.

We have chosen a symmetric device, where the vanishing of $g_2^{(0)}$
dictates that the differential conductance at low 
$V$ takes the form $g^{(0)}(V)=g_1^{(0)}+(g_3^{(0)}/2)V^2$.  
The $V$-symmetry observed in the inset of Fig.\ \ref{fig2} is broken once
a perturbing tip induces second-order corrections $g_2^{(1)}$.
The ``tilting'' of the differential conductance pattern appearing in 
the top panel is an effect of the tip-caused breaking of the spatial and bias 
symmetries, and can be directly confronted with experiments.
In the lower panel we present $g^{(1)}(V)$ (solid) and its linear 
contribution $g_1^{(1)}$ (dashed), corresponding to the indicated cuts at 
$V=0.05\Delta_1/e$ and $V=0$ in the upper panel, respectively.

The $\lambda_\mathrm{F}/2$-periodic oscillations characteristic of 
$g_1^{(1)}$ [\onlinecite{jalabert2010}, \onlinecite{gorini2013}] are also 
present in the nonlinear conductance corrections, with a phase shift building 
up at large distances. This phase shift could be related with the well-defined 
phase conditions observed as a function of $V$ in the difference of conductance 
changes between two tip positions.\citep{jura2009} 
Interestingly, while the oscillations decay as $(k_\mathrm{F}x)^{-1}$ for
$g_1^{(1)}$, there is no such decay for the leading nonlinear 
term $g_2^{(1)}$, which dominates the conductance correction at large distances.
The origin of this experimentally observable effect lies in the 
$V_{1(2)}$ dependence of the wave-vectors $k_{1(2)}$ and the 
$e^{2ik_{1(2)} x}$ terms present in the matrix elements of the tip-induced 
perturbation. Indeed, independently of the details of the model describing 
the system, if $g(\br_T)$ shows interference fringes,
then nonlinear corrections should become dominant away from the constriction.
This observation is in line with recent experimental results, 
\cite{sellier2013,sellier_private} where the oscillating tip-induced corrections 
at finite bias voltage do not decrease in magnitude with increasing tip-QPC 
distances when scanning in some regions of the 2DEG adjacent to the QPC.

This is unusual, since an increase of $V$ has a radically different effect
from that of a temperature rise. Starting from the linear regime, the 
temperature averaging effect would reduce the oscillations of $g_1^{(1)}$, while 
increasing $V$ leads to the dominance of $g_{2,3}^{(1)}$ with robust spatial 
oscillations.

\begin{figure}
\includegraphics[width=\linewidth]{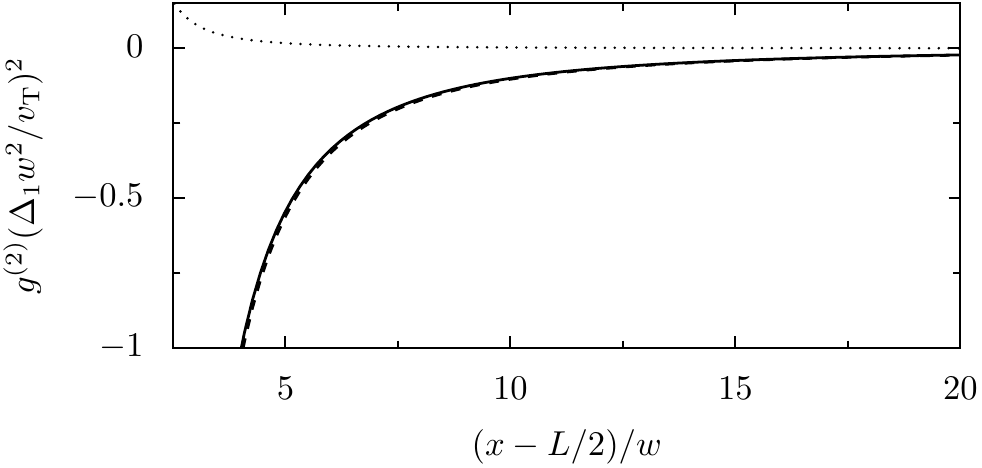}
\caption{\label{fig4}
The tip-induced change $g^{(2)}$ of the differential conductance up to 
third-order terms as in Fig.\ \ref{fig3}, but for the QPC tuned to the first 
conductance plateau (point ``p'' in Fig.\ \ref{fig2}), as a function of the 
tip position along the $x$-axis. 
Since $g^{(1)}$ is suppressed on plateaus, the lowest conductance correction 
is obtained to second order in the tip potential.
The dashed and solid line corresponds to voltages of $V=0$ and $eV=0.1\Delta_1$, 
respectively. The dotted line shows the first rectifying nonlinear contribution 
$g_2^{(2)}V$ for the second voltage value. 
}
\end{figure}
The above-discussed rectification effect is the most prominent
at conductance steps, and considerably reduced on a conductance plateau. 
The leading conductance corrections in the perturbative weak-probe limit for an 
abrupt QPC on the first conductance plateau are shown in Fig.\ \ref{fig4}. 
They are quadratic in $v_\mathrm{T}$ and dominated by the linear conductance 
correction $g_1^{(2)}$. The conductance corrections do not exhibit spatial 
oscillations, and the nonlinear rectifying contributions are relevant only for 
rather large values of $eV/\Delta_1$.

%%%%%%%%%%%%%%%%%%%%%%%%%%%%%%%%%%%%%%%%%%%%%%%%%%%%%%%%%%%%%%%%%%%%%%%%%%%%%%%%%
\section{Conclusions}
%%%%%%%%%%%%%%%%%%%%%%%%%%%%%%%%%%%%%%%%%%%%%%%%%%%%%%%%%%%%%%%%%%%%%%%%%%%%%%%%%

We have developed a gauge-invariant theory for the weak 
nonlinear effects of a nanostructure probed by Scanning Gate Microscopy. 
Despite working in the limit of a non-invasive probe, 
we have demonstrated that the tip can induce a nonlinear (rectifying) 
conductance in a geometrically symmetric device. 
We have quantified such an effect for the case of an abrupt QPC, 
showing it to be physically significant and experimentally attainable. 
At a conductance step the tip-generated lowest nonlinear transport coefficient 
$g_2$ shows $\lambda_\mathrm{F}/2$-periodic oscillations with an amplitude 
that does not decrease with the QPC-tip distance, as long as the transport 
remains phase-coherent.
In particular, such a phenomenon should be model-independent and appear 
whenever the SGM signal shows such tip-position dependent oscillations. 

The tip-induced oscillations between the 0.7 and the zero-bias anomalies 
observed in Ref.~\onlinecite{sellier2013} happen for bias voltages in the 
scale of $\mu V$.  Subtle many-body effects are beyond the scope of the present work, 
based on a one-particle approach yielding results on the scale 
of the constriction quantization energy (${\rm m}eV$).  
Understanding the consequences that an SGM tip has on the differential conductance 
at such scales is a necessary ingredient in the interpretation of the experimental 
results.

%%%%%%%%%%%%%%%%%%%%%%%%%%%%%%%%%%%%%%%%%%%%%%%%%%%%%%%%%%%%%%%%%%%%%%%%%%%%%%%%%%%%%
\acknowledgments

We are grateful to J.-L.\ Pichard for stimulating discussions. 
We thank B.\ Brun, K.\ Ensslin, T.\ Ihn, A.\ A.\ Kozikov, and H.\ Sellier 
for useful discussions and the communication of unpublished 
experimental results. 
Financial support from the French National Research Agency 
ANR (Project No.\ ANR-08-BLAN-0030-02), from the CEA (DSM-Energie-Meso-Therm),
from the German Research Foundation DFG (TRR80),
and from the European Union within the Initial Training Network NanoCTM 
is acknowledged.

\bibliography{sgm_biblio} % Produces the bibliography via BibTeX.

\end{document}